\documentclass[11pt]{article}
\usepackage{graphicx}
\usepackage[a4paper]{geometry}
\usepackage[latin1]{inputenc}
\usepackage[T1]{fontenc}
\usepackage{verbatim}
\usepackage{graphicx}
\usepackage{authblk}
\usepackage{amsmath}
\usepackage{amssymb}
\usepackage{nicefrac}
\usepackage{subfigure}
\usepackage{pgfplots}
\usepgfplotslibrary{dateplot}
\usepackage{multido}
\usepackage{calc}

\newtheorem{THEOREM}{Theorem}
\newenvironment{theorem}{\begin{THEOREM} \hspace{-.85em} {\bf .} }%
                        {\end{THEOREM}}
\newtheorem{LEMMA}{Lemma}
\newenvironment{lemma}{\begin{LEMMA} \hspace{-.85em} {\bf .} }%
                      {\end{LEMMA}}
\newtheorem{COROLLARY}{Corollary}
\newenvironment{corollary}{\begin{COROLLARY} \hspace{-.85em} {\bf .} }%
                          {\end{COROLLARY}}
\newtheorem{PROPOSITION}{Proposition}
\newenvironment{proposition}{\begin{PROPOSITION} \hspace{-.85em} {\bf .} }%
                            {\end{PROPOSITION}}
\newtheorem{PROPERTY}{\normalfont\textit{Property}}
\newenvironment{property}{\begin{PROPERTY} \hspace{-.85em} {\itshape .} \rm}%
                            {\end{PROPERTY}}
\newtheorem{DEFINITION}{Definition}
\newenvironment{definition}{\begin{DEFINITION} \hspace{-.85em} {\bf .} \rm}%
                            {\end{DEFINITION}}
\newtheorem{DEFINITIONL}[DEFINITION]{Definition}
\newenvironment{definitionl}{\begin{DEFINITIONL} \hspace{-.85em} {\bf .} \rm}%
                            {\end{DEFINITIONL}}
\newtheorem{EXAMPLE}{Example}
\newenvironment{example}{\begin{EXAMPLE} \hspace{-.85em} {\bf .} \rm}%
                            {\end{EXAMPLE}}
\newtheorem{ALGORITHM}{Algorithm}
\newenvironment{algorithm}{\begin{ALGORITHM} \hspace{-.85em} {\bf .} \rm}%
                          {\end{ALGORITHM}}
\newtheorem{PROCEDURE}{Procedure}
\newenvironment{procedure}{\begin{PROCEDURE} %
\setcounter{proclineno}{0}\hspace{-.85em} {\bf .} \rm}%
                          {\end{PROCEDURE}}
\newtheorem{REMARK}{Remark}
\newenvironment{remark}{\begin{REMARK} \hspace{-.85em} {\bf .} \rm}%
                            {\end{REMARK}}
\newtheorem{REMARKL}[REMARK]{Remark}
\newenvironment{remarkl}{\begin{REMARKL} \hspace{-.85em} {\bf .} \rm}%
                            {\end{REMARKL}}
\newtheorem{CLAIM}{Claim}
\newenvironment{claim}{\begin{CLAIM} \hspace{-.85em} {\bf .} \rm}%
                            {\end{CLAIM}}
\newtheorem{CLAIMEMPH}{Claim}
\newenvironment{claimemph}{\begin{CLAIMEMPH} \hspace{-.85em} {\bf .} }%
                            {\end{CLAIMEMPH}}
\newtheorem{HYPOTHESIS}{Hypothesis}
\newenvironment{hypothesis}{\begin{HYPOTHESIS} \hspace{-.85em} {\bf .}
    \rm}%
                            {\end{HYPOTHESIS}}
\newtheorem{HYPOTHESISL}[HYPOTHESIS]{Hypothesis}
\newenvironment{hypothesisl}{\begin{HYPOTHESISL} \hspace{-.85em} {\bf .} \rm}%
                            {\end{HYPOTHESISL}}
\newtheorem{FACT}{Fact}
\newenvironment{fact}{\begin{FACT} \hspace{-.85em} {\bf .} \rm}%
                            {\end{FACT}}
\newtheorem{FACTL}[FACT]{Fact}
\newenvironment{factl}{\begin{FACTL} \hspace{-.85em} {\bf .} \rm}%
                            {\end{FACTL}}
\newcommand{\thm}{\begin{theorem}}
\newcommand{\lem}{\begin{lemma}}
\newcommand{\pro}{\begin{proposition}}
\newcommand{\propw}{\begin{property}}
\newcommand{\prop}{\begin{property}}
\newcommand{\dfn}{\begin{definition}}
\newcommand{\dfnl}{\begin{definitionl}}
\newcommand{\rem}{\begin{remark}}
\newcommand{\reml}{\begin{remarkl}}
\newcommand{\clm}{\begin{claim}}
\newcommand{\clme}{\begin{claimemph}}
\newcommand{\hypt}{\begin{hypothesis}}
\newcommand{\hyptl}{\begin{hypothesisl}}
\newcommand{\fct}{\begin{fact}}
\newcommand{\fctl}{\begin{factl}}
\newcommand{\xam}{\begin{example}}
\newcommand{\alg}{\begin{algorithm}}
\newcommand{\proc}{\begin{procedure}}
\newcommand{\cor}{\begin{corollary}}
\newcommand{\prf}{\noindent{\bf Proof.} }

\newcommand{\ethm}{\end{theorem}}
\newcommand{\elem}{\end{lemma}}
\newcommand{\epro}{\end{proposition}}
\newcommand{\eprop}{\bbox\end{property}}
\newcommand{\epropw}{\end{property}}
\newcommand{\edfn}{\bbox\end{definition}}
\newcommand{\edfnl}{\end{definitionl}}
\newcommand{\erem}{\bbox\end{remark}}
\newcommand{\ereml}{\end{remarkl}}
\newcommand{\eclm}{\bbox\end{claim}}
\newcommand{\eclme}{\end{claimemph}}
\newcommand{\ehypt}{\bbox\end{hypothesis}}
\newcommand{\ehyptl}{\end{hypothesisl}}
\newcommand{\efct}{\bbox\end{fact}}
\newcommand{\efctl}{\end{factl}}
\newcommand{\exam}{\bbox\end{example}}
\newcommand{\ealg}{\end{algorithm}}
\newcommand{\eproc}{\end{procedure}}
\newcommand{\ecor}{\end{corollary}}
\newcommand{\eprf}{\bbox}

\newcommand{\bbox}{\vrule height7pt width4pt depth1pt}

\def\is{\textsc{max independent set}}
\def\np{\textbf{NP}}

\def\bb{{branch-\&-bound}}
\def\gnp{$\mathcal{G}(n,p)$}

\def\tnp{$\mathbb{T}(n,p)$}
\def\tnpd{\mathbb{T}(n,p)}

\def\enpd{\mathbb{E}}

\def\eps{\varepsilon}

\let\geq\geqslant
\let\ge\geqslant
\let\leq\leqslant
\let\le\leqslant

\makeatletter
\def\@fnsymbol#1{\ensuremath{\ifcase#1\or (a)\or (b)\or (c)\or (d)\or *\or \S \or
   \mathsection\or \mathparagraph\or \|\or **\or \dagger \or \ddagger \or \dagger\dagger
   \or \ddagger\ddagger \else\@ctrerr\fi}}
\makeatother

\def\Prb{\mathbb{P}}
\def\Esp{\mathbb{E}}

\usepackage{pgf,tikz}
\usetikzlibrary{shapes,snakes,patterns,arrows,automata,positioning,patterns}

\def\EE{\mathbb{E}}
\def\PP{\mathbb{P}}
\title{\textbf{Average-case complexity of a branch-and-bound algorithm for \is{}, under the~$\mathcal{G}(n,p)$ random model}}
\author[1]{N. Bourgeois}
\author[2]{R. Catellier\footnote{The work has been performed while author was with CEREMADE, Université Paris-Dauphine and CNRS UMR 7534}}
\author[3,4]{T. Denat}
\author[3,4]{V.~Th.~Paschos} 
\affil[1]{SAMM, Université Paris~I, \texttt{nbourgeo@phare.normalesup.org}}
\affil[2]{IRMAR-CHL, Université de Rennes, \texttt{remi.catellier@univ-rennes1.fr}}
\affil[3]{Paris Sciences et Lettres Research University, Universit\'e Paris-Dauphine}
\affil[4]{LAMSADE, CNRS, UMR~7243, \texttt{\{denat,paschos\}@lamsade.dauphine.fr}}

\begin{document}

\maketitle

\begin{abstract}
We study average-case complexity of branch-and-bound for \is{} in random graphs under the $\mathcal{G}(n,p)$ distribution. In this model every pair~$(u,v)$ of vertices belongs to~$E$ with probability~$p$ independently on the existence of any other edge.  We make a precise case analysis, providing phase transitions between subexponential and exponential complexities depending on the probability~$p$ of the random model.
\end{abstract}

\section{Introduction}\label{intro}

Given a graph~$G(V,E)$, the \is{} problem consists of determining a maximum-size subset~$V'$ of~$V$ such that, for any $(v_i,v_j) \in V' \times V'$, $(v_i,v_j) \notin E$. \is{} is one of the most known \np{}-hard problems (among the~21 ones of Karp's list~\cite{karpcl}) and furthermore, among the hardest problems to approximate in polynomial time, since it is inapproximable within approximation ratio~$n^{\epsilon-1}$, unless $\textbf{P} = \np{}$, for any $\epsilon > 0$~\cite{jzucker}. On the other hand, dealing with solution of \is{} by moderately exponential algorithms, i.e., algorithms computing an optimal solution within provably non-trivial (exponential) running time, a bunch of such algorithms have been developed during the last ten years, leading to complexities whose the basis of the exponential is now below~1.2, around~1.19~\cite{DBLP:journals/corr/XiaoN13}.

The goal of this paper is to study average complexity of branch-and-bound for \is{} in random graphs. Although mathematical tools for average-case analysis of algorithms exist since many years (to our knowledge,~\cite{knuth1} was the first book on these tools and~\cite{flajolet-sedgewickbook} the most recent one; in the meanwhile and after 2008 decades of papers have handled this topic), there are quite a few of results in average-case complexity of exact algorithms for \np{}-hard problems. 

For instance, \bb{} is perhaps the most popular and widely known technique for solving \np{}-hard problems. But, despite its popularity, no systematic study for obtaining non-trivial upper-bounds for  the complexity of the this method have been conducted until now, neither in the worst-case, nor in the average-case complexity framework. This is our objective in this paper.

\medskip

\noindent
\textbf{Our contribution.}  We study average complexity of \bb{} in random graphs under the \gnp{} random model~\cite{bo}. We make a precise case analysis, providing phase transitions between subexponential and exponential complexities depending on the probability~$p$ of the model. Our study concerns, two versions of \bb{}: the simplified one that consists of a simple exhaustive search, and a refined one where nodes of the search tree are provided with an evaluation function indicating, informally, for vertex~$v$, the best independent set that can be hoped in a leaf of the subtree rooted at~$v$.

\section{Preliminaries}\label{prelim}

Branch-and-bound algorithms are among the most common and popular strategies in combinatorial optimization for solving \np{}-hard problems used since the origins of combinatorial optimization. For a graph problem whose solutions are subsets of the vertex-set of the input-graph (this is the case of \is{}), a standard \bb{} implementation consists of subsequently handling vertices one by one and of examining two configurations: one where the vertex at hand is part of the solution and another one when the vertex is not part. In this way, \bb{} builds a fictitious binary tree, called \textit{search-tree}, internal nodes of which can be seen as partial solutions that will be expanded to global ones at the leaves of the tree. In other words, suppose that vertices are handled in the order~$v_1$ to~$v_n$. Then, level~$i$ of the tree, corresponds to vertex~$v_i$ and the nodes of level~$i$ correspond to all possible solutions that can be built if one takes, or does not take,~$v_i$ under any feasible configuration for the vertices~$v_1$ to~$v_{i-1}$. So, any partial solution at level~$i$ will be decomposed into two subsets: those containing vertex~$v_{i+1}$ and those that do not contain it. Search tree is pruned on an internal node~$t$ when either the partial solution represented by~$t$ is infeasible, or no hope exists that expansion of this solution will lead to an optimal solution for the whole problem. This can be done by providing internal nodes with an evaluation function (called \textit{evaluation}), that characterizes the ``potential'' of the corresponding partial solutions (for this reason, in what follows we will use the term potential instead of evaluation). Evaluation for node~$t$ gives a bound on the value of the best leaf of the subtree rooted at~$t$ (recall that leaves of the search tree correspond to solutions for the whole instance). For a maximization (resp., minimization) problem, the potential is an upper (resp., lower) bound of this value. The \bb{} tree is then pruned on some node when either the partial solution there, is not feasible, or the evaluation of the node is smaller (resp., greater) than the value of an already known solution. For more details about \bb{}, the interested reader can be referred to~\cite{ps}.

The goal of the paper is to study the average complexity of \bb{} in random graphs for \is{}. Two main random graph models have been studied in the literature, models~$\mathcal{G}(n,p)$ (also called random binomial model) and~$\mathcal{G}(n,m)$ (also called uniform random graph model)~\cite{bo}. In model~$\mathcal{G}(n,p)$, every pair~$(u,v)$ of vertices belongs to~$E$ with probability~$p$ (i.e., $\PP((u,v) \in E) = p$, $\forall (u,v) \in V\times V$),  independently on the existence of any other edge in~$G$. In model~$\mathcal{G}(n,m)$, any graph~$G$ on~$n$ vertices with~$m$ edges has the same occurrence-probability, in other words, $\forall G'(V,E')$, such that $|E'| = m$, $\PP(G=G') = {m \choose {n \choose 2}}$. 

We deal of two versions of \bb{}. The first one is the exhaustive search method. It can be seen as a \bb{} where the search-tree is pruned only on branches rooted at nodes that correspond to infeasible partial solutions. In the second version, nodes of the search-tree are provided with an evaluation function that, as described above, is an upper bound on the best solution that will be built when the node is completely unfolded.  For \is{} a simple evaluation for a node~$t$ of the \bb{} tree is function $u(S) = |S| + (n - \ell(t))$, where~$S$ denotes the partial solution already constructed at a node~$t$ of the search-tree,~$n$ denotes the order of the input-graph and~$\ell(t)$ is the level of~$t$ in the tree. Recall that at level~$\ell$, the \bb{} algorithm has handled~$\ell$ vertices of~$G$. Function~$u(S)$, also called ``evaluation of~$S$'', simply expresses the fact that if~$S$ is the partial independent set built when handling some node~$t$,~$\ell(t)$ vertices of~$G$ have been fixed. The ideal solution one can hope from~$t$ is the one that includes in~$S$ the whole set of the $n-\ell(t)$ unfixed vertices of~$G$. 
Recall that the evaluation of a leaf (since infeasible subtrees have already been discarded) is the size of the independent set corresponding to this leaf.

Close to the problematic of this paper is the one by~\cite{banderieretal_siam}, where a very nice analysis of the exhaustive search for \is{} is done, and phase transitions between exponential, subexponential, and polynomial average-case complexities under the~$\mathcal{G}(n,p)$ model are given. The main result there, is an asymptotic upper bound on the average running time (denoted by~$\mu_n$) of the exhaustive search when applied on~\gnp{} random graphs (formula~(1.4) in~\cite{banderieretal_siam}). It is shown that when $p \gg \nicefrac{\log^2{n}}{n}$,~$\mu_n$ becomes subexponential.
%

\medskip

\noindent
\textbf{Our results.} Dealing with exhaustive search, we first generalize the the result by~\cite{banderieretal_siam} showing that this algorithm runs in subexponential average time whenever $n\cdot p \rightarrow \infty$. Then we show that, when~$n\cdot p$ is a fixed constant~$k$, the average complexity of exhaustive search remains exponential and in this case we give explicit upper and lower bounds for this complexity and show that both of them tend to~1, when $k \rightarrow \infty$. More importantly, we study average complexity of a ``real'' \bb{} method, where potentials are associated with the nodes of the search-tree. In this case we show the following:
\begin{itemize}
\item average complexity of \bb{} is subexponential as far as $p \neq \nicefrac{k}{n}$ for any fixed constant~$k$; this means, roughly, that \bb{} remains subexponential for both large and very small values of~$p$ and constitutes a major difference of the method with respect to the simple exhaustive search algorithm;
\item on the other hand, if $p = \nicefrac{k}{n}$, we show that the \bb{} remains exponential in average; for this case, we prove that its average running time is bounded above by~$O(1.867^n)$.
\end{itemize} 
%
%
%
\textbf{Notations.} For the rest of the paper, $\mathbb{T}(n,p)$ will denote the average complexity of \bb{} in a binomial random graph~$G(V,E)$ with parameters~$(n,p)$;~$\enpd[\# S(G)]$ will denote the expectation of the number of independent sets in a binomial random graph~$G$;~$P(t,i)$ the partial solution formed by vertices in the set $\{v_1,\ldots,v_i\}$ on node~$t$ of the \bb{}-tree (recall that vertices are handled in the order $v_1, v_2, \ldots, v_n$);~$G_i$ will denote the graph~$G[\{v_1,\ldots,v_i\}]$ (induced by the vertex-set $\{v_1,\ldots,v_i\}$); finally,~$u(S)$ will denote the potential of a partial solution~$S$; if $S = P(t,i)$, then $u(S) = |V(S)| + n - i$.

\section{Exhaustive search: pruning by infeasibility}\label{simpleb&b}

\subsection{Upper bounds for~\tnp}

%

Consider  a random $(n,p)$-binomial graph~$G$. 
By definition of the $(n,p)$-binomial random model, the probability that a set of~$k$ vertices is an independent set of~$G$ is equal to $(1-p)^{{k \choose 2}}$; henceforth:
\begin{equation}\label{es}
\mathbb{E}[\# S(G)] = \sum_{i=0}^n{n \choose i}(1-p)^{{i \choose 2}} \leqslant n \max\limits_{i \leqslant n}\left\{{n \choose i}(1-p)^{{i \choose 2}} \right\}
\end{equation}
%
Note that the number of independent sets in the subgraph induced by the~$k$ first fixed vertices is strictly greater than the number of independent sets induced in the $k-1$ first such vertices (since the~$k$-th vertex is an independent set by itself); so:
\begin{equation}\label{boundcn}
\tnpd{} \leqslant n\cdot\enpd[\# S(G)] = n \sum_{i=0}^n{n \choose i}(1-p)^{{i \choose 2}} \leqslant n^2 \max\limits_{i \leqslant n}\left\{{n \choose i}(1-p)^{{i \choose 2}} \right\}
\end{equation}
Using Stirling's formula, by some very elementary algebra, the following holds for the last term of~(\ref{boundcn}):
\begin{equation}\label{boundbinomial}
{n \choose i}(1-p)^{{i \choose 2}} = O\left(\left(\frac{en}{i}\right)^i(1-p)^{\nicefrac{i^2}{2}}\right)
\end{equation}
\begin{lemma}\label{pfixed}
If~$p$ is a fixed constant, then~\tnp{} is subexponential with~$n$ and bounded above by~$O(n^{\ln{n}})$.
\end{lemma}
\prf
Note that $(\nicefrac{en}{i})^i\cdot(1-p)^{\nicefrac{i^2}{2}} = e^{i(1+\ln{n}-\ln{i})+\ln(1-p)\nicefrac{i^2}{2}} \leqslant e^{i(\ln{n})+\ln(1-p)\nicefrac{i^2}{2}}$. The derivative of function $f(i) = i(\ln{n})+\ln(1-p)\nicefrac{i^2}{2}$ is decreasing with~$i$ and admits a maximum when $i = \nicefrac{\ln{n}}{(-\ln(1-p))}$. So, $e^{\ln{n}+\ln(1-p)(\nicefrac{i^2}{2})} \leqslant \nicefrac{n^{\ln{n}}}{(-2\ln(1-p))}$. Since~$p$ is supposed to be fixed, $-2\ln(1-p)$ is a fixed constant. So, combination of~(\ref{boundcn}) and~(\ref{boundbinomial}) leads to $\tnpd = O(n^{\ln{n}})$, as claimed.~\eprf
\lem\label{ptendssl20}
When $p = \nicefrac{\phi(n)}{n}$ for some function~$\phi$ such that $\phi(n) = o(n)$, $\phi \rightarrow \infty$ when $n \rightarrow \infty$,~\tnp{} is subexponential with~$n$ and satisfies $\tnpd = O(e^{n\cdot(\nicefrac{(2\ln\phi(n)+\ln^2\phi(n))}{2\phi(n)})})$.
\elem
\prf
Revisit function $f(i) = i\ln{n}+\ln(1-p)\frac{i^2}{2}$. Its derivative is $f'(i) = \ln{n}-\ln{i} + i\ln(1-p)$. When $p \rightarrow 0$, $\ln(1-p) = -p +o(p)$; so, $f'(i) = \ln(\nicefrac{n}{i}) - ip + ip\epsilon(p)$, with $\epsilon(p) \rightarrow 0$. Function~$f$ being convex, its maximum is attained when $f'(i) = 0$, i.e., when:
\begin{eqnarray*}
\ln{n} - \ln{i} - ip = ip\epsilon(p) &\Longrightarrow& \ln{n} = \ln\left(ie^{ip(1+\epsilon(p))}\right) \;\; \Longrightarrow \;\; n = ie^{ip(1+\epsilon(p))} \\
&\Longrightarrow& pn =  ipe^{ip(1+\epsilon(p))} = ie^{ip} + p\varepsilon(p) \\
&\Longrightarrow& (n-\varepsilon(p)) = ie^{ip} \;\; \Longrightarrow \;\; i = \frac{W[p(n-\varepsilon(p))]}{p}
\end{eqnarray*}
where~$W(x)$ denotes Lampert's function defined by $W(x)  =x e^{W(x)}$.
So, $i = \nicefrac{W(pn)}{p}$ and ${n \choose i}(1-p)^{{i \choose 2}} \leqslant e^{\nicefrac{(2W(np)+W^2(np))}{2p}} + O(1)$. Since $p = \nicefrac{\phi(n)}{n}$, using~(\ref{boundcn}), we have:
$$
e^{\frac{2W(\phi(n))+W^2(\phi(n))}{2\phi(n)}} \leqslant e^{\frac{2\ln(\phi(n))+\ln^2(\phi(n))}{2\phi(n)}} \longrightarrow 0
$$
and in this case~\tnp{} is as claimed.~\eprf
\lem\label{ptendsqu20}
When $p = \nicefrac{k}{n}$, for a fixed constant~$k$,~\tnp{} is exponential in~$n$.
\elem
\prf
From the discussion in the proof of Lemma~\ref{ptendssl20}, $\tnpd = O(e^{n\cdot(\nicefrac{2W(np)+W^2(np)}{2p})})$, and for $p = \nicefrac{k}{n}$, this gives $\tnpd = O(e^{n\cdot(\nicefrac{2W(k)+W^2(k)}{2k})})$, with $g(k) = \nicefrac{2W(k)+W^2(k)}{2k}$ being constant when~$k$ is constant.~\eprf

Note, furthermore, that function $g(k) = \nicefrac{2W(k)+W^2(k)}{2k}$ decreases with~$k$. For instance, $g(1.25) = 1.995$ and $g(6.15) = 1.5$. 

Lemmata~\ref{pfixed}, \ref{ptendssl20} and~\ref{ptendsqu20} derive the following theorem.
\begin{theorem}\label{exsearchthm}
The average running time of exhaustive search for \is{} when running in $(n,p)$-binomial random graph~$G$ of order~$n$ is:
\begin{enumerate}
\item\label{ena} subexponential when~$p$ is constant, or $p = \nicefrac{\phi(n)}{n}$ for some function~$\phi$ such that $\phi(n) = o(n)$, with $\phi \rightarrow \infty$ when $n \rightarrow \infty$;
\item exponential for $p = \nicefrac{k}{n}$, if~$k$ is a fixed constant.
\end{enumerate}
\end{theorem}

\subsection{Lower bounds}

Recall that any feasible independent set can be associated with a leaf of the branching tree built by the the \bb{}-algorithm. In other words, any such set have been explored by the algorithm. Thus, $\mathbb{T}(n,p) \geqslant \mathbb{E}[\# S(G)]$. Revisit~(\ref{es}); when $p\cdot n \rightarrow 0$, we have:
\begin{equation}\label{inf1}
\binom{n}{i} (1-p)^{\binom{i}{2}} \geq e^{i(\ln n - \ln i)+\binom{i}{2} \ln(1-p)} = e^{i(\ln n - \ln i)+\binom{i}{2} \ln(1-p)} \backsim e^{i(\ln n - \ln i)-p\binom{i}{2}}
\end{equation}
The maximum value for~(\ref{inf1}) is reached when $\nicefrac{\partial [i(\ln n - \ln i)) -p\binom{i}{2}]}{\partial i}=0$, i.e., when:
$$
\ln\left(\frac{n}{i}\right) - 1-p\cdot i+\frac{p}{2} =0 \Rightarrow i=\frac{W(e^{\frac{p}{2}-1} np)}{p}
$$
In this case:
$$
\binom{n}{i} (1-p)^{\binom{i}{2}} \geq e^{\frac{2W(e^{\frac{p}{2}-1}np)+W^2(e^{\frac{p}{2}-1}np)}{2p}} \longrightarrow e^{\frac{2W(e^{-1}np)+W^2(e^{-1}np)}{2p}}
$$
and $\tnpd \geqslant n\cdot e^{\frac{2W(e^{-1}np)+W^2(e^{-1}np)}{2p}}$.


On the other hand, when $p = \nicefrac{k}{n}$, $k \in \mathbb{R}$:
\begin{eqnarray}\label{inf2}
f(n\cdot p) = f(k) &=& \binom{n}{i} (1-p)^{\binom{i}{2}} \;\; \geq \;\; e^{\frac{2W(e^{\frac{p}{2}-1}np)+W^2(e^{\frac{p}{2}-1}np)}{2p}} \nonumber \\
&=& e^{\frac{2W(e^{-1}np)+W^2(e^{-1}np)}{2p}} \;\; = \;\; e^{\frac{2W(e^{-1}k)+W^2(e^{-1}k)}{2k}}
\end{eqnarray}
Function~$f(k)$ in~(\ref{inf2}) decreases with~$k$ and $f(0) \rightarrow e^{\nicefrac{1}{e}}$. So, for $p = \nicefrac{k}{n}$, $\tnpd \geqslant e^{\nicefrac{n}{e}}$.

\section{Associating evaluations with the nodes of the branch-and-bound tree}\label{b&b}

In this section we analyze the average complexity of a \bb{} algorithm with potentials on the nodes of the search tree and following a best-first rule. This rule implies that the node further developed is that of the largest potential (evaluation). Observe that the root of the \bb{} tree has potential~$n$ and that \textit{any ``child''~$P_0$ of a partial solution~$P$ has $u(P_0) \leqslant u(P)$}. Hence, if we have a partial solution~$P_{i+1}$ handled at step $i+1$ of the algorithm, then~$P_{i+1}$ is either a partial solution already available in step~$i$ (thus with $u(P_{i+1}) \leqslant u(P_i)$), or a child of~$P_i$ (the solution handled at step~$i$). Let~$S^*$ be the first global solution built by the method; obviously $u(S^*) = |S^*|$. Then, any partial solution non-handled yet have some potential that is at most~$|S^*|$ (otherwise they would have been handled before). So,~$S^*$ is a maximum independent set and any other branch of the search-tree will be pruned.

So, during the unravelling of the algorithm, a partial solution~$P(t,i)$ at node~$t$ is expanded only if its vertices are independent and its potential is at least equal to the cardinality of~$S^*$. Thus:
\begin{eqnarray}\label{tnp1}
\tnpd &\leq& \sum\limits_{i=0}^{n}\sum\limits_{P(t,i) \subseteq V\left(G_i\right)}\PP\left((P(t,i) \text{ is an independent set } )\cap \left(n-i+ |P(t,i)| \geq \left|S^*\right|\right)\right) \nonumber \\
& =& \sum\limits_{i=0}^{n}\sum\limits_{P(t,i) \subseteq V\left(G_i\right)}\PP(P(t,i) \text{ is an independent set }) \nonumber \\
& & \mbox{} \cdot \PP\left(n-i+ |P(t,i)| \geq \left|S^*\right|\; |\; P(t,i) \text{ is an independent set }\right) 
\end{eqnarray}

\subsection{The basic theorem}

We are ready now to prove the following basic theorem that will be used in what follows.
\begin{theorem}\label{basicthm}
Given a graph~$G(V,E)$ with $|V| = n$ and $|E| = m$, the average complexity of \bb{} with potentials associated with the nodes of the search tree when running in~$G$ is bounded above by:
$$
\tnpd \leqslant \sum\limits_{i=0}^{n}\sum\limits_{k=0}^{i} \binom{i}{k} (1-p)^{\binom{k}{2}}\cdot\PP\left(m \geq \frac{n^2}{2\cdot\left(n-i +k\right)}-\frac{n}{2}\right)
$$
\end{theorem}
\prf
Revisit~(\ref{tnp1}) and observe that, if the size of an independent set~$S$ of~$G$ (obviously $|S| \leqslant |S^*|$) is already known, it holds that:
\begin{eqnarray}\label{tnp2}
& & \sum\limits_{i=0}^{n}\sum\limits_{P(t,i) \subseteq V\left(G_i\right)}\PP(P(t,i) \text{ is an independent set }) \nonumber \\
& & \mbox{} \cdot \PP\left(n-i+ |P(t,i)| \geq \left|S^*\right|\; |\; P(t,i) \text{ is an independent set }\right) \nonumber \\
&\leqslant& \sum\limits_{i=0}^{n}\sum\limits_{P(t,i) \subseteq V\left(G_i\right)}\PP(P(t,i) \text{ is an independent set }) \nonumber \\
& & \mbox{} \cdot \PP\left(n-i+ |P(t,i)| \geq |S|\; |\; P(t,i) \text{ is an independent set }\right)
\end{eqnarray}
By Turan's Theorem~\cite{tu}, any maximal for inclusion independent set of a graph~$G$ of order~$n$ has size at least~$\nicefrac{n}{\bar{d}+1}$, where~$\bar{d}$ is the average degree of~$G$. Denoting by~$m$ the number of edges of~$G$, it holds that $m = \nicefrac{n\cdot\bar{d}}{2}$ and, for any~$k$:
\begin{equation}\label{turan}
\PP\left(\frac{n}{\bar{d}+1} \geqslant n - i + k\right) = \PP\left(m \geqslant \frac{n^2}{2\cdot(n-i+k)} - \frac{n}{2}\right)
\end{equation}
In order to complete the proof of the theorem, we need the following lemma.
\lem\label{bblem1}
For any binomial random graph~$G(V,E)$, for any $\eta \in [0,n]$ and for any subset~$V'$ of~$V$:
$$
\PP\left(m > \eta \; | \; V' \text{ independent set }\right) \leqslant \PP(m > \eta)
$$
\elem
\prf[\textit{Lemma~\ref{bblem1}}]
Consider a set $V' \subseteq V$ and partition~$E$ into subsets $E' = \{(x,y) \in E: x \notin V' \vee y \notin V'\}$ and $E'' = \{(x,y) \in E: x \in V' \wedge y \in V'\}$. Obviously,~$V'$ is an independent set iff $E'' = \emptyset$. On the other hand, $\PP(m > \eta) = \PP(|E'|+|E''| > \eta)$. So:
\begin{eqnarray}\label{secondline}
\PP\left(\left|E'\right| + \left|E''\right| >\eta \; | \; \left|E''\right| =0\right) &=& \frac{\PP\left(\left(\left|E'\right| + \left|E''\right| >\eta\right) \cap \left(\left|E''\right| =0\right)\right)}{\PP\left(\left|E''\right| = 0\right)} \nonumber \\
&=& \frac{\PP\left(\left(\left|E'\right| >\eta\right) \cap \left(\left|E''\right| =0\right)\right)}{\PP\left(\left|E''\right| = 0\right)} \nonumber \\
&=& \frac{\PP\left(\left|E'\right| >\eta\right) \cdot \PP\left(\left(\left|E''\right| =0\right)\right)}{\PP\left(\left|E''\right| = 0\right)} \nonumber \\
&=& \PP\left(\left|E'\right| >\eta\right) 
\end{eqnarray}
expression~(\ref{secondline}) holding because,~$E'$ and~$E''$ are independent (edges in~E are Bernoulli i.i.d. variables).

To conclude, it suffices to take into account that~$|E'|$ follows a binomial law:
$$
\mathcal{B}\left(\binom{n}{2}-\binom{\left|V\left(G'\right)\right|}{2},p\right)
$$
whereas~$m$ follows a binomial law:
$$
\mathcal{B}\left(\binom{n}{2},p\right)
$$
We so have: 
$$
\PP\left(\left|E'\right| > \eta\right) \leqslant \PP(m > \eta) \Longrightarrow  \PP(m > \eta)  \geqslant \PP\left(m > \eta \; | \; V' \text{ independent set }\right)
$$
The proof of Lemma~\ref{bblem1} is now completed.~\eprf

In order to complete the proof of the theorem, it suffices to put together~(\ref{tnp1}), (\ref{turan}) and Lemma~\ref{bblem1}.~\eprf

In what follows, we separate the possible values of~$p$ in the following three categories: 
\begin{itemize}
\item $n\times p \rightarrow \infty$;
\item $n\times p \rightarrow 0$;
\item $n\times p = k>0$
\end{itemize}
In the case of $n\times p \rightarrow \infty$ the exhaustive search is already subexponential, by Item~\ref{ena} of Theorem~\ref{exsearchthm}. Since \bb{} with potentials is faster than exhaustive search, this result also holds for the former.

\subsection{Average complexity when $n\times p \rightarrow 0$}


As~$m$ follows a binomial law, we will strongly use some concentration inequalities due to~\cite{chernoff1952,hoeffding} giving exact deviation from the expectation.  In what follows, we set $\phi(\delta)=(1+\delta)\ln(1+\delta)-\delta$.
\begin{lemma}\label{concentrationlem} [Chernoff concentration inequality]
Suppose $X_1,\dots,X_N$ are Bernoulli i.i.d. random variables, set $S_N=\sum_{i=1}^N X_i$ and $\mu=\EE[S_N]$. Then for any $\delta>0$ the following bound holds:
\[\PP\left(S_N\ge \mu(1+\delta)\right)\le e^{-\mu \phi(\delta))}\]
\end{lemma}
\prf
Let~$b$ be the parameter of Bernoulli, and recall that $\mu=pN$. For all $\lambda>0$, we have:
\[\PP\left(S_N\ge \mu(1+\delta)\right)=\PP\left(e^{\lambda S_N}\ge e^{\lambda \mu(1+\delta)}\right)\] 
 We then use the Markov inequality; recall that the~$X_i$'s are i.i.d and $\EE[e^{\lambda X_1}]=1+p(e^\lambda-1)$. We have:
 \begin{align*}
 \PP\left(S_N\ge \mu(1+\delta)\right)
 &\le e^{-\lambda N p (1+\delta)}\left(1+p(e^\lambda-1)\right)^N \le e^{-N\left(\lambda p (1+\delta)- p(e^\lambda-1) \right)} \\
 & = e^{-\mu\left(\lambda (1+\delta)+ (e^\lambda-1)\right)}
 \end{align*}
 where the second inequality comes from the fact that $\ln\left(1+p (e^\lambda-1)\right)\le p(e^\lambda-1)$.
 Finally, for the last bound, we set $\lambda=\ln(1+\delta)$, and we have the result.~\eprf


In what follows, we will use the notation $\gamma_n$ to denote a generic function with at most polynomial growth. This function may vary from lines to lines. 

If we decompose the complexity by increasing potential of vertex, we have:
\[\mathbb{T}(n,p) =\sum_{x=0}^n w_n(u)\]
where~$u$ is a potential-value and $w_n(u)$ is the expectation of the number of nodes in the search-tree with potential equal to~$u$. In what follows, we shall write~$p_n$ instead of~$p$ to underline the fact that $p \to_n 0$. It holds that:
\[w_n(u)=\sum_{i=n-u}^{n}\binom{i}{i-(n-u)}\left(1-p_n\right)^{\binom{i-(n-u)}{2}}\PP\left(m\ge \frac{n^2}{2u}-\frac{n}{2}\right)\]
Thereby, we can argue that:
\begin{align}\label{eq:bound_non_optimal}
w_n(u) &\le 2^n \PP\left(m\ge \frac{n^2}{2u}-\frac{n}{2}\right) \nonumber \\
w_n(u) &\le\sum_{i=n-u}^{n}\binom{i}{i-(n-u)}(1-p_n)^{\binom{i-(n-u)}{2}}
\end{align}
Remark that~$m$ follows a Binomial law of parameter $\binom{n}{2}$ and $p_n$. 
We can now decompose the complexity $\mathbb{T}(n,p)$ in two different sums, with different behaviors. Let $C_n\in [0,1]$, we can write
$$
T_{n,p}  = \sum_{u=0}^{nC_n}w_n(u) +\sum_{u=nC_n}^{n}w_n(u) := \mathbb{T}^1(n,p) +\mathbb{T}^2(n,p)
$$

\subsubsection{Small potentials}

Set $f_n = p \cdot n$. Then, for the small potentials the following bound holds.
\begin{proposition}
Let $\phi(\delta)=(1+\delta)\ln (1+\delta)-\delta$ and set:
\begin{align*}
\delta_n &=\phi^{-1}\left(\frac{2\ln 2}{f_n}\right) \\
C_n &=\frac{1}{1+\left(1+\delta_n\right)\frac{n-1}{n}f_n}
\end{align*}
Then $\mathbb{T}^1(n,p)$ has at most polynomial growth. 
\end{proposition}
\prf
Revisit~\eqref{t1+t2} and observe that in the term dealing with~$\mathbb{T}^1(n,p)$,~$u$ is less  than~$n\cdot C_n$;  hence, using~\eqref{eq:bound_non_optimal}, we have:
 \begin{equation}\label{t1+t2}
\mathbb{T}^1(n,p) \le \gamma_n 2^n \PP\left(m\ge \frac{n}{2}\left(\frac1{C_n}-1\right)\right)
 \end{equation}
Furthermore, by the definition of~$C_n$ we have:
 \[\PP\left(m\ge \frac{n}{2}\left(\frac1{C_n}-1\right)\right)= \PP\left(m\ge \binom{n}{2}p_n\left(1+\delta_n\right)\right)\]
 Observe that $\EE[m]=\binom{n}{2}p_n$, so:
 \begin{equation*}
 \mathbb{T}^1(n,p) \le \gamma_n 2^n \PP\left(m\ge \EE[m]\left(1+\delta_n\right)\right)
 \end{equation*}
 We are now able to apply Chernoff bound getting:
	\begin{equation*}
 \mathbb{T}^1(n,p) \le \gamma_n 2^n e^{-\binom{n}{2}p_n\phi\left(\delta_n\right)} \leqslant \gamma_n e^{-\frac{n}{2}\left( f(n)\phi(\delta_n) - 2\ln 2\right)}
 \end{equation*}
As $\phi(\delta_n)=\nicefrac{2\ln 2}{f(n)}$ the result follows.~\eprf
\rem
In fact we can chose $\delta_n \ge \phi^{-1}(\nicefrac{2 \ln 2}{f_n})$ and as~$\phi$ is non decreasing  the result will be still true, with a better bound on~$\mathbb{T}^1(n,p)$.~\erem 
\rem
 When $p_n=\nicefrac{k}{n}$ we have:
\[ C_n = \frac{1}{1+k\frac{n-1}{n}+\phi^{-1}\left(\frac{2\ln 2}{k}\right) k\frac{n-1}{n}}\]
Then,  the second term is needed in order to use the concentration Lemma~\ref{concentrationlem}.~\erem

\paragraph{Study of $\delta_n$ and $C_n$ when $f_n\to 0$.} For this case, one can prove the following
\begin{proposition}
When $f_n\to 0$, then:
\begin{align*}
\delta_n &= \frac{2\ln 2)}{f_n \ln\left(\frac{1}{f_n}\right)}+o\left(\frac{1}{f_n \ln\left(\frac{1}{f_n}\right)}\right) \\
C_n &= 1-\ln\left(\frac1{f_n}\right)^{-1}+o\left(\ln\left(\frac1{f_n}\right)^{-1}\right) 
\end{align*}
\end{proposition}
\prf
Set:
\[
H(x)=\exp\left\{W\left(\frac{x-1}{e}\right)+1\right\}-1=\frac{x-1}{W(\frac{x-1}{e})}-1
\]
where~$W(x)$ is the Lampert function.  Then, an easy computation shows that $\delta_n=H(\nicefrac{2\ln 2}{f_n})$.
 
Furthermore, when $x\to \infty$, $W(x)= \ln x+o(\ln x)$. Then: 
$$
H(x) = \frac{x-1}{W\left(\frac{x-1}{e}\right)}+1 = \frac{x-1}{\ln\left(\frac{x-1}{e}\right)+o(\ln x)}+1 =\frac{x}{\ln x}+o\left(\frac{x}{\ln x}\right)
$$
Since $f_n\to 0$:
\[
\delta_n=\frac{2\ln 2}{f_n \ln\left(\frac{2\ln 2}{f_n}\right)}+o\left(\frac{2\ln 2}{f_n \ln\left(\frac{2\ln 2}{f_n}\right)}\right)
\]
as claimed.
 
We are going now to compute the asymptotic for~$C_n$. Recall that $f_n=o(\ln(\nicefrac{1}{f_n}))$. We then have:
 \begin{align*}
 C_n&=\left(1+(1+\delta_n)(n-1)\frac{f_n}{n}\right)^{-1} = \left(1+(f_n+f_n \delta_n)(1+o(1))\right)^{-1} \\
 &=\left(1+\left(\ln\left(\frac1{f_n}\right)^{-1}+o\left(\ln\left(\frac1{f_n}\right)^{-1}\right)\right)(1+o(1))\right)^{-1} \\
 &=\left(1+\ln\left(\frac1{f_n}\right)^{-1}+o\left(\ln\left(\frac1{f_n}\right)^{-1}\right)\right)^{-1} =1-\ln\left(\frac1{f_n}\right)^{-1}+o\left(\ln\left(\frac1{f_n}\right)^{-1}\right) 
 \end{align*}
and the proof of the proposition is completed.~\eprf

\subsubsection{Large potentials, case $f_n\to 0$}

We now handle the part of $\mathbb{T}(n,p)$ dealing with large potentials. In this case the following proposition holds.
\begin{proposition}
 When $f(n)\to 0$,~$\mathbb{T}^2(n,p)$ has subexponetial growth. Furthermore, there exists a constant $h>0$ such that, 
 \[
 \mathbb{T}^2(n,p) \le \gamma_n \exp\left(h\cdot n\cdot\left(\ln\frac{1}{f_n}\right)^{-1}\ln\left(\ln\frac{1}{f_n}\right)\right)
 \] 
\end{proposition}
\prf
By the definition of~$\mathbb{T}^2(n,p)$, it holds that:
\begin{align*}
 \mathbb{T}^2(n,p) & = \sum_{u=nC_n+1}^n \sum_{i=n-u}^n\binom{i}{i-(n-u)}(1-p_n)^{\binom{i-(n-u)}{2}}  \le \sum_{u=nC_n+1}^n \sum_{i=n-u}^n\binom{i}{i-(n-x)} \\
 &\le \sum_{u=nC_n+1}^n \sum_{i=n-u}^n\binom{i}{n-u}
 \end{align*}
Furthermore, the~$\sup$ in~$i$ is reached when $i=n$, hence:
\[
\mathbb{T}^2(n,p)\le \gamma_n \sum_{u=nC_n+1}^n\binom{n}{n-u} = \sum_{u=0}^{n(1-C_n)}\binom{n}{u}
\]
Finally, as far as $1-C_n<\nicefrac{1}{2}$, the following bound holds:
\begin{equation*}
 \mathbb{T}^2(n,p)\le \gamma_n e^{n\cdot H\left(1-C_n\right)}
 \end{equation*}
 where $H(x)=-x\ln x - (1-x)\cdot\ln(1-x)$~(\cite{FG06} lemma 16.19).

We will now study the asymptotic convergence of~$\mathbb{T}^2(n,p)$. Recall that:
 \[
 C_n = 1-\ln\left(\frac1{f_n}\right)^{-1}+o\left(\ln\left(\frac1{f_n}\right)^{-1}\right) 
 \]
 We set $\eps_n=\ln(\nicefrac1{f_n})^{-1}$. Then:
 \begin{align*}
 -\left(1-C_n\right)\ln\left(1-C_n\right) &= -\left(\eps_n+o\left(\eps_n\right)\right)\left(\ln\left(\eps_n\right)+o\left(\ln\left(\eps_n\right)\right)\right) = -\eps_n \ln \eps_n+o\left(\eps_n \ln \eps_n\right) \\
-C_n\ln\left(C_n\right) & = -(1+o(1))\ln\left(1-\eps_n+o\left(\eps_n\right)\right) =\eps_n + o\left(\eps_n\right) 
 \end{align*}
and finally:
\[
H\left(1-C_n\right) \sim -\eps_n \ln \eps_n 
\]
It is easy to see now that here exists a constant $h>0$ such that $H(1-C_n)>-h\eps_n \ln\eps_n$.~\eprf

The discussion above proves the following theorem.
\begin{theorem}
The average case complexity of \bb{} with potentials on the nodes of the serach-tree when running in a random binomial graph is subexponential when $n\times p \to 0$.
\end{theorem}

\subsection{Average complexity when $n\times p = k>0$}

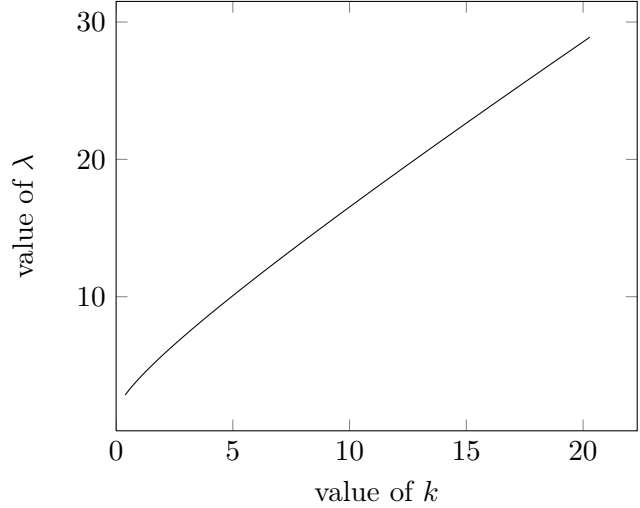
\begin{figure}[h!]
\begin{center}
\begin{tikzpicture}
\pgfplotstableread{./lambda.data}
\datatable
\begin{axis}[xlabel = value of $k$, xmin = 0, ylabel = value of $\lambda$]
\addplot [
color = black,
mark = none,
]table [y index = 1] from \datatable;
\end{axis}
\end{tikzpicture}
\caption{The curve for~$\lambda$. Indeed, $\lambda$ is very close to $(\nicefrac{4(k+1)}{3})+1$.}\label{curvelambda}
\end{center}
\end{figure}

We now handle the case where edge-probabilities are equal to~$\nicefrac{k}{n}$, for some $k \in \mathbb{R}$. We first prove the following proposition.
\begin{proposition}
Let $p = \nicefrac{k}{n}$. There exists some~$\lambda$ such that, for any $u \leq \nicefrac{n}{\lambda}$,~$w_n(u)$ is polynomial.
More precisely,~$\lambda$ is the solution of:
\begin{equation}\label{zsol}
2\ln 2-(k+1) + \lambda + (\lambda-1)\ln\frac{k}{\lambda-1}
\end{equation}
Or equivalently:
\begin{eqnarray*}
\lambda &=& 1 + k\left(1+\phi^{-1}\left(\frac{2}{k}\ln 2\right)\right)\\
\phi(z) &=& (1+z)\ln(1+z)-z
\end{eqnarray*}
\end{proposition}
\prf
Set $N = \nicefrac{n}{2}\cdot(n-1)$ and $K = (\nicefrac{n}{2})\cdot(\nicefrac{n}{u}-1)$ and recall that $k=np$ and that:
\begin{equation}\label{fbound}
\mathbb{T}^1(n,p)(u) \leq 2^n\Prb\left(m\geq K\right)
\end{equation}
Also notice that:
\begin{equation*}
\Esp(m) = \frac{n}{2}(n-1)p = \frac{k(n-1)}{2}
\end{equation*}
We restrict ourselves to the case $K > \Esp(m)$; otherwise there is no hope for~(\ref{fbound}) to be a polynomial bound. This is true whenever
$u \leq \nicefrac{n}{(k+1)}$.

Now, since we are above the expectation, we get:
\begin{align*}
\mathbb{T}^1(n,p) &\leq 2^nn^2\Prb\left(m =  K\right) \leq 2^nn^2\left(\frac{k}{n}\right)^K\left(1-\frac{k}{n}\right)^{N-K}{N \choose K} \\
\leq& n^4\exp\left\{\ln 2n + K\ln\frac{k}{n} + (N-K)\ln\left(1-\frac{k}{n}\right) \right. \\
& \mbox{} \left. - K\ln\frac{K}{N}-(N-K)\ln\left(1-\frac{K}{N}\right)\right\} \leq n^4e^{\frac{n}{2}\nu(u)}
\end{align*}
with
\begin{align*}
\nu(u) &= 2\ln 2 + \left(\frac{n}{u}-1\right)\ln k - \left(\frac{n}{u}-1\right)\ln n -k\left(1-\frac{1}{u}\right) - \left(\frac{n}{u}-1\right)\ln\left(\frac{n}{u}-1\right) \\
&  \mbox{} + \left(\frac{n}{u}-1\right)\ln(n-1)-\left(n-\frac{n}{u}\right)\ln\left(n-\frac{n}{u}\right)+\left(n-\frac{n}{u}\right)\ln(n-1) \\
&= 2\ln 2 + (n-1)\ln\left(1-\frac{1}{n}\right) + \left(\frac{n}{u}-1\right)\ln k - k\left(1-\frac{1}{u}\right) \\
&  \mbox{} - \left(\frac{n}{u}-1\right)\ln\left(\frac{n}{u}-1\right)-\left(n-\frac{n}{u}\right)\ln\left(1-\frac{1}{u}\right) \\
&= 2\ln 2 - 1 +\left(\frac{n}{u}-1\right)\ln k - k - \left(\frac{n}{u}-1\right)\ln\left(\frac{n}{u}-1\right)+\frac{n}{u} + O\left(\frac{n}{u^2}\right) \\
&= \mu(\lambda) + O\left(\frac{n}{u^2}\right)\\
\end{align*}
with
\begin{equation*}
\mu(\lambda) = 2\ln 2 + \lambda - (k+1) + (\lambda-1)\ln\frac{k}{\lambda-1}
\end{equation*}
In other words,~$w_1(u)$ is polynomial if $\mu(\lambda)\leq 0$.

\begin{figure}
\begin{center}
\begin{tikzpicture}
\pgfplotstableread{./cpx.txt}
\datatable
\begin{axis}[xlabel = value of $k$, xmin = 0, xmax =15, ymin=1, width=12cm, height = 6cm, ylabel = complexity $\gamma$]
\addplot [
color = black,
mark = none,
]table [y index = 1] from \datatable;
\end{axis}
\end{tikzpicture}
\caption{The curve for~$\gamma(k)$.}\label{gammacurve}
\end{center}
\end{figure}
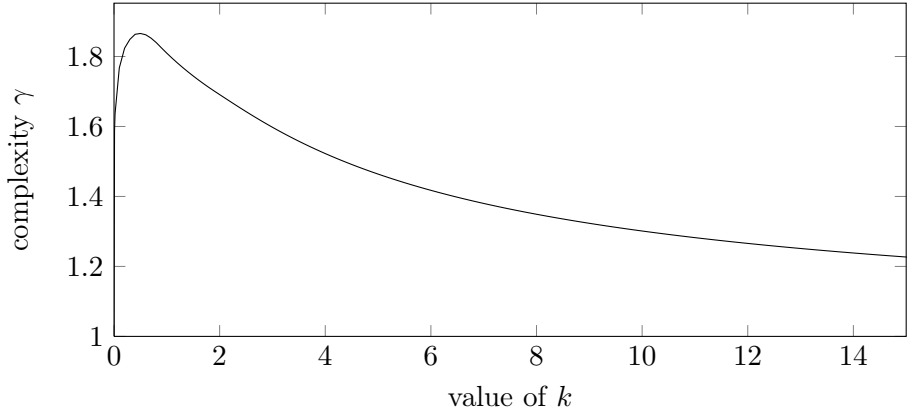

\begin{center}
\begin{table} 
$$
\begin{array}{l|ccccccc}
k & 10^{-5} & 10^{-4} & 10^{-3} & 10^{-2} & 10^2 & 10^3 & 10^4 \\
\hline
\gamma &  1.0004 & 1.399 & 1.473 & 1.589 & 1.052 & 1.008 & 1.001 \\
\end{array}
$$
\caption{Extremal values of $\gamma$.}\label{gammatable}
\end{table}
\end{center}

Note that $\mu' = \ln(\nicefrac{k}{(\lambda-1)})$ which is nonegative according to the hypothesis above, so there is exactly one solution to~(\ref{zsol}). Figure~\ref{curvelambda} gives an approximation of the curve for~$\lambda$.~\eprf
\begin{theorem}
If $p=\nicefrac{k}{n}$, then the running time is bounded with $\gamma(k)^n$ where (see also Figure~\ref{gammacurve} and Table~\ref{gammatable}):
\begin{itemize}
\item $\gamma \leq \gamma(0.47) \leq 1.867$;
\item $\gamma \rightarrow 1$, when $k\to 0$;
\item $\gamma \rightarrow 1$ when $k\to \infty$.
\end{itemize}
\end{theorem}
\prf
Fix $s = i-n+u$. Then:
\begin{equation*}
\mathbb{T}^2(n,p) \leq \sum_{u=0}^n \sum_{i = n-u}^n w_n(u) \leq \sum_{s=0}^n \sum_{i = s}^{n} w_n\left(s+n-i\right) \leq (n+1)^2 \max_{s=0}^n \left\{\max_{i = s}^{n}\left\{e^{\phi_n(i,s)}\right\}\right\}
\end{equation*}
with:
\begin{align*}
\phi_n(i,s) &= s\ln \frac{i}{s} + (i-s)\ln \frac{i}{n-s} - \frac{k\cdot s^2}{2n}\\
&= i\ln i - s\ln s - (i-s) \ln(i-s) - \frac{k\cdot s^2}{2n}
\end{align*}
It holds that:
\begin{equation*}
\frac{\partial \phi_n}{\partial i} = (1+\ln i) - (1+\ln(i-s)) = \ln\frac{i}{i-s} > 0
\end{equation*}
Hence, the maximum is reached when $n_0=n$ and thus $s=u$ which leads to:
\begin{equation*}
\mathbb{T}^2(n,p) \leq (n+1)^2 \max_{s=0}^n\left\{e^{\phi_n(n,s)} \right\}= (n+1)^2 \max_{u=0}^n\left\{e^{\psi_n(u)}\right\}
\end{equation*}
with: 
\begin{equation}
\label{psi}
\psi_n(u) = n\ln n - u\ln u - (n-u) \ln(n-u) - \frac{k\cdot u^2}{2n}
\end{equation}
For any given~$k$, the running time is bounded above by:
\begin{equation*}
\gamma(k)^n \leq \max_u\left\{\min\left\{\mathbb{T}^1(n,p),\mathbb{T}^2(n,p)\right\}(u)\right\} \leq \max_u\left\{\exp\left\{\min\left\{\psi_n(u), \nicefrac{\mu(u)n}{2}\right\}\right\}\right\}
\end{equation*}
We already know that~$\mu$ is increasing. On the other hand, things are a little bit more complicated for~$\psi_n$:
\begin{equation*}
\psi_n'(s) = -(1+\ln s) + (1+\ln (n-s)) -\frac{ks}{n} = \ln (\frac{n-s}{s}) - \frac{k\cdot s}{n}
\end{equation*}
This last function is clearly a (decreasing) bijection from $]0,n[$ onto $\mathbb{R}$. Thus, there exists one single~$s_0$ such that $\psi_n(s_0)$ is a maximum.

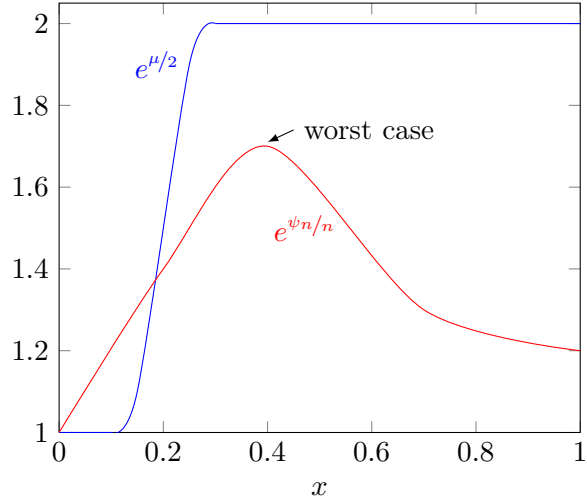
\begin{figure} 
\begin{center}
\begin{tikzpicture}
\begin{axis}[xmin =0,ymin=1,xmax=1,ymax=2.05,xlabel=$x$]
\addplot[mark= none, color=blue] coordinates {
            (0,1)
            (0.1,1)
        };
\addplot[smooth, mark= none, color=blue] coordinates {
            (0.1,1)
            (0.15,1.1)
            (0.25,1.9)
            (0.3,2)
        };
\addplot[mark= none, color=blue] coordinates {
            (0.3,2)
            (1,2)
        };
\addplot[smooth, mark= none, color=red] coordinates {
			(0,1)            
            (0.2,1.4)
            (0.4,1.7)
            (0.7,1.3)
            (1,1.2)
        };
\draw[latex-] (40,71) -- +(5,3) node[right] {worst case};
\draw (25,90) node[left,blue] {$e^{\nicefrac{\mu}{2}}$}; 
\draw (55,50) node[left,red] {$e^{\nicefrac{\psi_n}{n}}$};
\end{axis}

\end{tikzpicture}
\caption{Here, $\Delta >0$ and thus $\mathcal{M} =\mathbb{T}^2(n,p)(s_0)$.}\label{schemabelow1}
\end{center}
\end{figure}

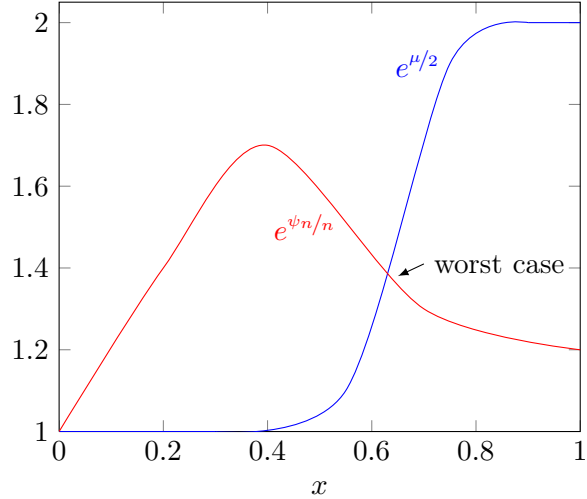
\begin{figure}[h!]
\begin{center} 
\begin{tikzpicture}
\begin{axis}[xmin =0,ymin=1,xmax=1,ymax=2.05,xlabel=$x$]
\addplot[mark= none, color=blue] coordinates {
            (0,1)
            (0.3,1)
        };
\addplot[smooth, mark= none, color=blue] coordinates {
            (0.3,1)
            (0.55,1.1)
            (0.75,1.9)
            (0.9,2)
        };
\addplot[mark= none, color=blue] coordinates {
            (0.9,2)
            (1,2)
        };
\addplot[smooth, mark= none, color=red] coordinates {
			(0,1)            
            (0.2,1.4)
            (0.4,1.7)
            (0.7,1.3)
            (1,1.2)
        };
\draw[latex-] (65,38) -- +(5,3) node[right] {worst case};
\draw (75,90) node[left,blue] {$e^{\nicefrac{\mu}{2}}$}; 
\draw (55,50) node[left,red] {$e^{\nicefrac{\psi_n}{n}}$};
\end{axis}

\end{tikzpicture}
\caption{Here, $\Delta < 0$ and thus $\mathcal{M} =\mathbb{T}^1(n,p)(\arg(\mathbb{T}^1(n,p)=\mathbb{T}^2(n,p)))$.}\label{schemabelow2}
\end{center}
\end{figure}

Now, there are two different cases depending on the sign of $\Delta = ((\nicefrac{\psi_n}{n}) - (\nicefrac{\mu}{2}))(s_0)$ as illustrated in Figures~\ref{schemabelow1} and~\ref{schemabelow2}.

We now have enough information to compute numerically the value of
$$
\mathcal{M} = \max_u\left\{\min\left\{\mathbb{T}^1(n,p),\mathbb{T}^2(n,p)\right\}\right\}
$$
for any~$k$. The asymptotic bounds directly derive from the following inequalities. When $k\to \infty$:
$$
\mathcal{M} \leq \max_u\left\{\mathbb{T}^2(n,p)\right\} \leq \max_u\left\{e^{n\left(1+\ln\frac{u}{n}-k\left(\frac{u}{n}\right)^2\right)}\right\} \leq \max_x\left\{e^{n\left(\frac{u}{n}-k\left(\frac{u}{n}\right)^2\right)}\right\} = e^{o(1)n}
$$
On the other hand, when $k \longrightarrow 0$ and for $u_0 = \nicefrac{n}{(1+k(1+\phi^{-1}\cot((\nicefrac{2}{k}) \cdot \ln 2)))}$ we have:
\begin{equation*}
\mathcal{M}_{u<u_0} \leq \max_{u<u_0}\left\{\mathbb{T}^1(n,p)\right\} = e^{o(1)n}
\end{equation*}
Furthermore:
\begin{align*}
\mathcal{M}_{u>u_0} &\leq \max_{u>u_0}\left\{\mathbb{T}^2(n,p)\right\} \\
 \leq& \max_{u>u_0}\left\{{n \choose n-u} \right\}\leq \max_{u>u_0}\left\{e^{n(1+\ln(1-(\nicefrac{u}{n})))}\right\} \leq e^{n(1-\frac{1}{k+1})} = e^{o(1)n}
\end{align*}
Thus we can conclude that with the appropriate value for~$p$, the average complexity cannot be higher that~$1.867^n$.~\eprf


\begin{thebibliography}{10}

\bibitem{banderieretal_siam}
C.~Banderier, H.~Hwang, V.~Ravelomanana, and V.~Zacharovas.
\newblock Analysis of an exhaustive search algorithm in random graphs and the
  $n^{c\log n}$-asymptotics.
\newblock {\em SIAM J.~Disc. Math.}, 28(1):342--371, 2014.

\bibitem{bo}
B.~Bollob\'{a}s.
\newblock {\em Random graphs}.
\newblock Academic Press, London, 1985.

\bibitem{chernoff1952}
H.~Chernoff.
\newblock A measure of asymptotic efficiency for tests of a hypothesis based on
  the sum of observations.
\newblock {\em Ann. Math. Statist.}, 23(4):493--507, 12 1952.

\bibitem{flajolet-sedgewickbook}
Ph. Flajolet and R.~Sedgewick.
\newblock {\em Analytic combinatorics}.
\newblock Cambridge University Press, 2008.

\bibitem{FG06}
J.~Flum and M.~Grohe.
\newblock {\em Parameterized complexity theory}.
\newblock Texts in The\-o\-re\-ti\-cal Computer Science. Springer-Verlag, 2006.

\bibitem{hoeffding}
W.~Hoeffding.
\newblock Probability inequalities for sums of bounded random variables.
\newblock {\em J. of the American Stat. Assoc.}, 58(301):13--30, 1963.

\bibitem{karpcl}
R.~M. Karp.
\newblock Reducibility among combinatorial problems.
\newblock In R.~E. Miller and J.~W. Thatcher, editors, {\em Complexity of
  computer computations}, pages 85--103. Plenum Press, New York, 1972.

\bibitem{knuth1}
D.~E. Knuth.
\newblock {\em The art of computer programming: fundamental algorithms},
  volume~1.
\newblock Addison-Wesley, Reading MA, 1969.

\bibitem{ps}
C.~H. Papadimitriou and K.~Steiglitz.
\newblock {\em Combinatorial optimization: algorithms and complexity}.
\newblock Prentice Hall, New Jersey, 1981.

\bibitem{tu}
P.~Tur\'an.
\newblock On an extremal problem in graph theory (in {H}ungarian).
\newblock {\em Mat.\ Fiz.\ Lapok}, 48:436--452, 1941.

\bibitem{DBLP:journals/corr/XiaoN13}
M.~Xiao and H.~Nagamochi.
\newblock Exact algorithms for maximum independent set.
\newblock {\em CoRR}, abs/1312.6260, 2013.

\bibitem{jzucker}
D.~Zuckerman.
\newblock Linear degree extractors and the inapproximability of max clique and
  chromatic number.
\newblock {\em Theory of Computing}, 3(6):103--128, 2007.

\end{thebibliography}

\end{document}